\documentclass[runningheads]{llncs}
\usepackage[T1]{fontenc}
\usepackage{amsmath,amsfonts}

\usepackage{array}
\usepackage{subfigure}
\usepackage{multirow}
\usepackage{booktabs}
\usepackage{stfloats}
\usepackage{url}
\usepackage{verbatim}
\usepackage{graphicx}
\usepackage{algorithm}
\usepackage{algpseudocode}
\usepackage{epstopdf}

\begin{document}
\title{Shavette: Safe Undervolting of Neural Network Accelerators via Algorithm-level Error Detection}
\titlerunning{Shavette: Safe Undervolting of Neural Network Accelerators}
\author{Mikael Rinkinen\inst{2} \and Lauri Koskinen\inst{2} \and Olli Silvén\inst{1} \and Mehdi Safarpour\inst{1}}

\authorrunning{M. Rinkinen et al.}
\institute{Center for Machine Vision and Signal Analysis, University of Oulu, Finland 
 \and
 Circuit and Systems Unit, University of Oulu, Finland
\\
\email{\{firstname.lastname\}@oulu.fi}}

\maketitle

\begin{abstract}
With transistor technology reaching its limits, exploring alternative approaches to enhance the energy efficiency of Deep Neural Network (DNN) accelerators has become essential. Operating at lower voltages substantially enhances energy efficiency. However, this approach is limited by increased sensitivity to Process, Voltage, and Temperature (PVT) variations, complicating the identification of the optimal energy-efficient voltage-frequency operating point, that is lowest voltage and highest clock frequency. Traditional approaches to adaptive voltage scaling significantly increase design complexity and are unsuitable for already fabricated devices. This paper introduces a novel application of algorithmic fault detection methods to enable reliable and aggressive voltage down-scaling in DNN accelerators. This work is the first to leverage algorithmic fault detection techniques specifically for reducing operational voltage in DNN accelerators as a practical, low-cost energy-saving solution. To showcase the solution, two popular DNN models, i.e., LeNet and VGG-16, were developed from scratch in C++/OpenCL to integrate the algorithmic error protection mechanisms. Experiments on a commercial AMD GPU demonstrated $18\%$ to $25\%$ energy savings without any accuracy loss and trivial, i.e., less than $4\%$, throughput compromise.


\end{abstract}

\section{Introduction}
Deep Neural Networks (DNNs) are becoming increasingly popular for consumer applications, including 5G connected edge devices such as autonomous vehicles \cite{tarkoma2023ai}. The reliance on DNNs in these devices presents a growing challenge. Their computational demands are substantially higher than conventional embedded applications, hence, DNNs strongly impact energy efficiency, straining battery life and impose severe thermal constraints. Already being at the edge of transistor technology, i.e., due to end of Moore's law and Dennard scaling, utility of alternative energy optimization approaches becomes more prominent.

One such approach involves operating at reduced voltages \cite{krishna2022global}. Lowering voltage quadratically reduces power consumption, e.g., a mere 20\% voltage reduction can cut the energy consumption to half. In the vicinity of the threshold voltages of the transistors, $V_{th}$, energy efficiency gains of 10x to 100x are provisioned \cite{ernst2003razor}. Unfortunately, as the operating voltage is reduced, sensitivity to Process, Voltage and Temperature (PVT) variations exacerbates, rendering commonly used Static Timing Analysis (STA) methods unreliable. 

Voltage reduction necessarily requires lowering frequency to meet timings \cite{ernst2003razor}. However, manufacturers often include a safety margin, typically around 10\% to 20\%, in their recommended voltage, implying the device might function with the same performance at even lower voltages. This margin accounts for potential variations in temperature, voltage, and process conditions. Studies have found different margin values: 10\% in commercial CPUs \cite{krishna2022global}, 20\% in Field-Programmable Gate Arrays (FPGAs) \cite{safarpour2021high}, and around 10\% in GPUs \cite{thomas2016emergpu}. If those voltage margins are eliminated safely, substantial power savings without throughput lost, i.e., frequency down-scaling, is achievable \cite{thomas2016emergpu}. It is difficult to remove margins in a manner that ensures no timing errors in logic circuits or bit-flip in memory systems are induced. The margin size varies due to PVT variations from die to die and as a function of time for the same chip, i.e., due to aging. STA does not provide reliable operating points either, hence, modeling the processor for optimal operating point, that is the lowest working voltage and highest clock frequency, becomes very challenging. GPUs in particular experience larger voltage variation due to IR drop, which needs to be considered \cite{thomas2016emergpu}.

Traditionally, voltage margin trimming techniques, e.g., Razor \cite{ernst2003razor}, rely on adding extra hardware into the digital circuit. Those were not conventionally supported by Computer Aided Design (CAD) tools and add substantial design complexity and incur significant costs \cite{hiienkari20200}. In contrast, this contribution introduces a straightforward, low-overhead solution\footnote{Source codes associated with this paper are available online at \url{https://github.com/NortHund/OCDNN}.}
to enable reliable elimination of voltage margins in DNN accelerators. 

To the best knowledge of authors, this is the first work on leveraging fault detection techniques for enabling a low-cost fully software aggressive undervolting solution. The work demonstrates the solution using real-world DNN model and processing platform. The results show substantial savings.  

Unlike earlier voltage down-scaling approaches, e.g., Razor \cite{ernst2003razor}, that require extensive intrusion into the hardware micro-architecture, ``Shavette'' only requires software changes and by borrowing concepts from fault tolerance community, makes safe and low overhead voltage down-scaling a possibility.



\section{Background and Motivation}

\begin{figure}[t!]
  \begin{center}
  \includegraphics[width=3.1in]{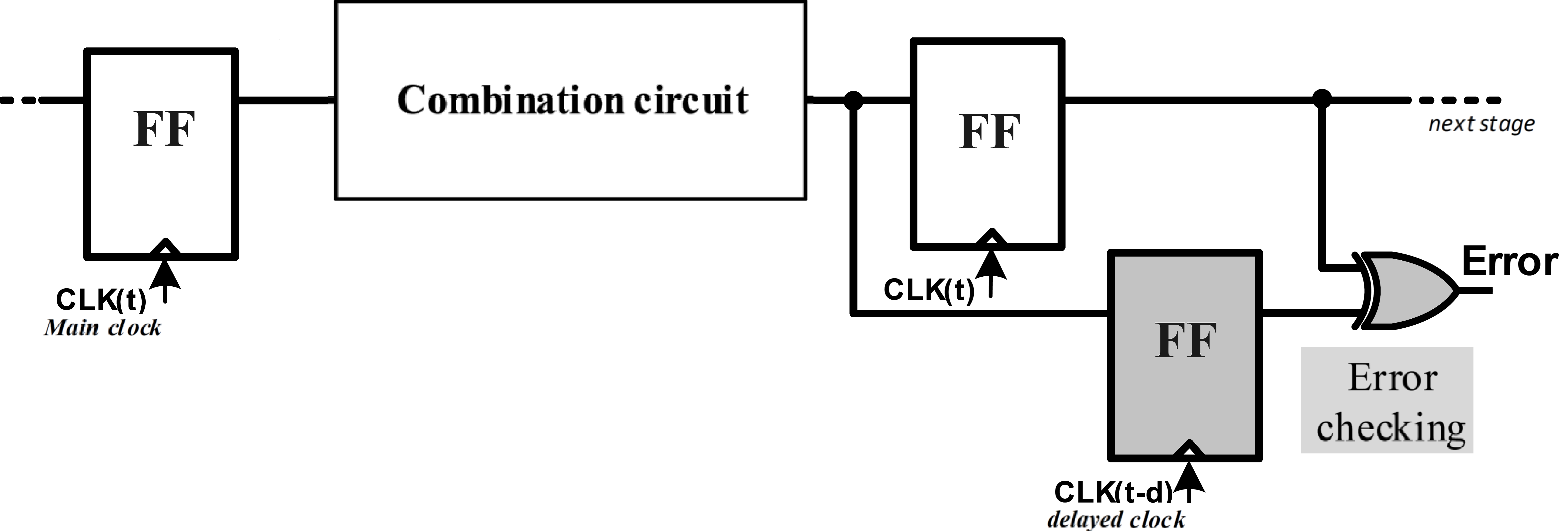}
  \vspace{-5pt}
  \caption{Simplified concept of Timing-Error-Detection circuit \cite{safarpour2021algorithm}.}\label{TED}
  \end{center}
  \vspace{-10pt}
\end{figure}

Two of the most investigated hardware based techniques for enabling reduced voltage operation are briefly discussed here. A simple option is to use adaptive delay monitoring circuits, e.g., via ring oscillators or delay chains.  This solution offers wider applicability but suffers from unreliability and does not ensure optimal voltage-frequency estimates \cite{kim2016adaptive}. The second approach is based on inserting Timing Error Detection (TED) based systems \cite{ernst2003razor}, like Error Detection Sequences (EDS), as is depicted in Fig.~\ref{TED}, into the critical paths of the circuit to monitor for any timing errors in-situ. As shown in Fig.~\ref{TED}, a set of extra secondary flip-flops clocked with a slight delay compared to the primary set of flip-flops are inserted the end of critical combinational logic paths. By detecting a late arrival signal, possibly due to reduced voltage operation, the timing errors become detectable. The detections from TED system are fed back into the voltage regulator to adjust the voltage such that minimum voltage is selected while no errors are observed. Although TED based solutions, e.g., the well-known Razor \cite{ernst2003razor}, enable significant voltage reduction, they come with high development complexities and overheads as those techniques cause significant hiccups in semiconductor design and manufacturing flow, e.g., by requiring extensive cell library characterization or secondary clock tree insertion. Finally, the hardware based techniques are only implementable in pre-silicon stage \cite{do2023evaluating}.

\section{Proposed Solution: Algorithmic Error Detection}
\label{proposed_solution}

As mentioned earlier, for enabling error-free computations at reduced voltages, a run-time error detection mechanism is needed. As an alternative to hardware based methods for enabling low voltage operation, the proposed solution relies on Algorithm Based Fault Tolerance (ABFT) which was originally introduced by Huang and Abraham \cite{huang1984algorithm} in 80s for detection and correction of errors in large matrix computations \cite{ensieh2025scissors}.

\subsection{Detecting error in linear layers}


Linear operations of DNNs comprise matrix multiplications and convolutions.
The original ABFT scheme \cite{huang1984algorithm} detects errors in matrix operations by adding checksum property in input matrices and then verifies correctness of computations by inspecting the checksum property in output matrices. A slightly different version of the the original ABFT scheme was employed for convolutional layers and fully-connected layers.

The models were ran without batching inputs, meaning that the fully-connected layers were 1-dimensional vectors multiplied with weight matrices. The weight matrices were modified to include an addition sum column and the resulting vector would hold a checksum value containing the sum of the other output values, as depicted in Eq.~\ref{mm_1}. 

Initially, for integration of ABFT, methods derived from referenced works \cite{marty2020safe} and \cite{zhao2020ft} were considered. While the approach of \cite{marty2020safe} theoretically appeared more appealing due to lower computational overhead, during later optimization phases, we found that a simplified variant from \cite{zhao2020ft} offered greater suitability for architectures with numerous cores, such as GPUs, and hence was implemented accordingly. 

The employed ABFT variant for convolutions utilizes a property of 3-dimensional convolutional layers, which is described by Eq.~\ref{ft_eq} and Eq.~\ref{ft_1}. In the equations, the output tensor is denoted by $O$, bias vector by $B$, input tensor by $D$ and the weight tensor by $W$. The convolution operation is denoted by $\otimes$. According to this property, an input checksum can be generated by calculating a convolution operation using the original input and associated weights- and biases sums, where values in the dimension $m$ (in Eq.~\ref{ft_1}) are added together. An output checksum can be generated by adding values of the original output together in the dimension $m$. The input and output checksums are equal 2-dimensional matrices, as described in Eq.~\ref{ft_2}. If these two checksums don't match, it is assumed that there has been an error in the computations of the layer. 

\begin{figure}[t!]
  \begin{center}
  \includegraphics[width=3.4in]{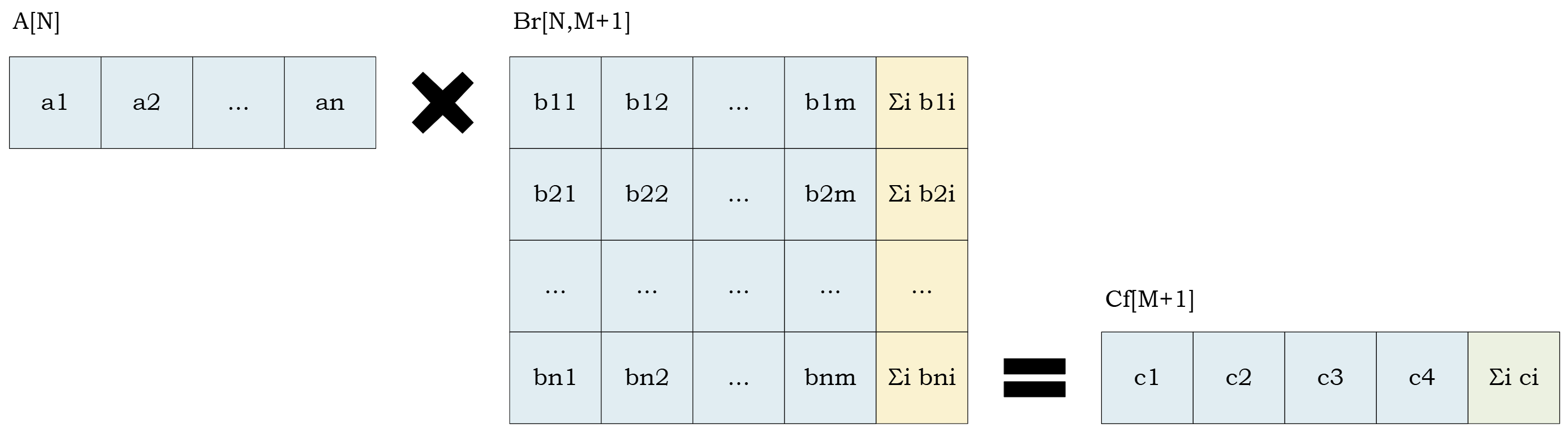}
  \vspace{-5pt}
  \caption{ABFT scheme for vector in matrix multiplications.}  \label{ABFT_for_mm}
  \end{center}
  \vspace{-5pt}
\end{figure}

\begin{figure}[t!]
  \begin{center}
  \includegraphics[width=3.4in]{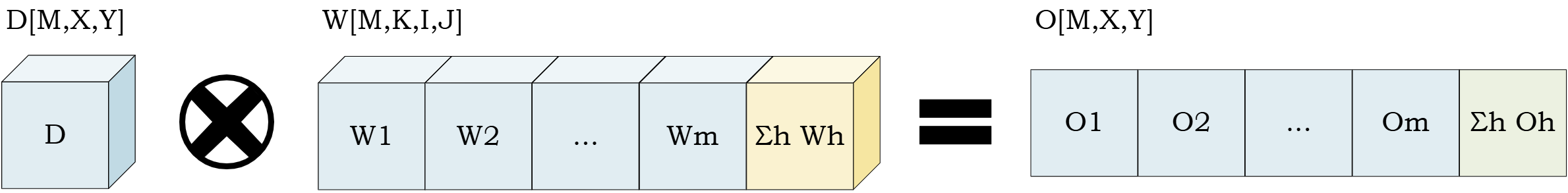}
  \vspace{-5pt}
  \caption{ABFT scheme for convolutions.}  \label{ABFT_for_conv}
  \end{center}
  \vspace{-10pt}
\end{figure}

The applied ABFT schemes are depicted in figures \ref{ABFT_for_mm} and \ref{ABFT_for_conv}.


\begin{equation}
    \sum_{m=0}^{M - 1} C_{m} = A \times \sum_{m=0}^{M - 1} B_m
    \label{mm_1}
\end{equation}

\begin{equation}
\begin{split}
    &O[m][x][y] = B[m] + 
    \sum_{k=0}^{Ch - 1} \sum_{i=0}^{R - 1} \sum_{j=0}^{R - 1}
    D[k][Ux+1][Uy+j] \times W[m][k][i][j] \\
    &0 \leq 0 \leq m < M, 0 \leq x, y < E, E = \frac{H-R+U}{U}
    \end{split}
    \label{ft_eq}
\end{equation}

\begin{equation}
    D \otimes W_1 + D \otimes W_2 = D \otimes (W_1 + W_2)
    \label{ft_1}
\end{equation}

\begin{equation}
    \sum_{m=0}^{M - 1} O_{m} = \sum_{m=0}^{M - 1} B_m + D \otimes \sum_{m=0}^{M - 1} W_m
    \label{ft_2}
\end{equation}

\subsection{Detecting errors in non-linear layers}
The ABFT schemes only work for linear operations while DNNs consist of non-linear layers, e.g., pooling and activation functions \cite{lecun1998gradient}, Double Redundant Module (DMR) approach was employed for those layers. Although DMR incurs large overheads, i.e. 100\%, but fortunately, computations of non-linear layers contribute to just of a few percent of total computations \cite{lecun1998gradient}. Hence, error detection through DMR adds tolerable overheads, when is restricted to only non-linear layers. Note, the redundant module must be implemented uncorrelated to the original one, e.g., with different instruction set. This is to avoid correlated errors. In experiments it was observed that errors always appear in linear operations before non-linear operations possibly due to exercising longer delay paths. Nonetheless DMR was still added to protect all layers against errors \cite{nia2007generalized}. 

It is important to highlight that, although ABFT and DMR have traditionally been used for fault tolerance, to the best of the authors' knowledge, this is the first instance of their application specifically aimed at reducing voltage operation for power savings in DNN processing.

\section{Experiments and Results}
\label{implementaiton_section}

\begin{algorithm}[t!]
\caption{ABFT-DMR Error Handling for Low Voltage GPU based Inference}
\begin{algorithmic}[1]
\State \textbf{Offline:} Precompute weight checksums for conv./FC layers
\State \textbf{Online:} CPU sends input data and model to GPU
\State GPU processes data layer-by-layer:
\begin{itemize}
\item Compute ABFT checksums for conv./FC layers
\item Compute DMR outputs for nonlinear layers
\item Generate prediction results
\end{itemize}
\State GPU returns checksums, DMR outputs, and prediction to CPU
\If{checksums and DMR outputs match expectations}
\State CPU accepts prediction
\Else
\State CPU rejects prediction; adjusts GPU voltage/frequency
\State Repeat inference (from step 2)
\EndIf
\end{algorithmic}
\label{Algorthim_1}
\end{algorithm}

As Algorithm~\ref{Algorthim_1} describes, in our implementation the inference of ABFT augmented DNN model is carried out within the undervolted accelerator, i.e., GPU (AMD RX 5600 XT), while its voltage is controlled by the CPU (Intel i5 3570k). The CPU act as the host and oversees the GPU computation by validating checksums and DMR outputs to detect and handle errors efficiently. Clock frequencies and voltage levels were adjusted using ``amdgpu-clocks'' tool \cite{amdgpu_clocks,MikaelsThesis}.

The inference output, i.e., DNN's prediction, along with checksums and DMR results are sent back to the CPU. The CPU checks for computational errors by verifying the correctness of linear and non-linear operations within the model as described in previous subsections. If no errors are observed the inference results from the neural network model is accepted, otherwise, e.g., the voltage/frequency is adjusted and inference operation is repeated. 

In this manner, the voltage of the GPU as the DNN accelerator, is dynamically adjusted to minimum error free operating voltage, without incurring errors in computations, neither degrading network accuracy nor causing system crash by timely detection of the Point of First Failure (PoFF). No additional on-chip or off-chip modifications are involved and hence the solution is truly applicable to the commercial off-the-shelf components \cite{MikaelsThesis}. To detect errors in linear parts, i.e., the convolutional and fully-connected layers, the convolution and matrix operations kernels were modified. Details of implementation are available in our online Git repository mentioned earlier. With non-linear layers the DMR modules used two different codes to avoid the same faults appearing in both, causing incorrect voting results. 

ABFT checksum for each layer was calculated on-the-fly within the  GPU. The checksums of weights were pre-computed for each model to minimize overheads. For inference, pre-computing the input checksums for weights is possible, however, training obviously requires updating the weights and hence re-computing the wight checksums.

The scheme in our experimentation was to keep the clock frequency and reduce the operating voltage until errors start appearing. Once errors detected the voltage can be retracted to slightly higher voltage. However, we continued reducing voltage to detect both the PoFF and point of crash  to assess the error detection capability. PoFF voltage turned out to be much higher than crash voltage, hence, the solution alarms voltage governor, way before approaching the crash point. In all experiments, the on-chip temperature sensor report was fluctuating around $55^{\circ}C$. Notice at each voltage step same test batch was used for estimating the accuracy.
\vspace{-7pt}

\begin{figure}[t!]
  \begin{center}
\includegraphics[width=3.9in]{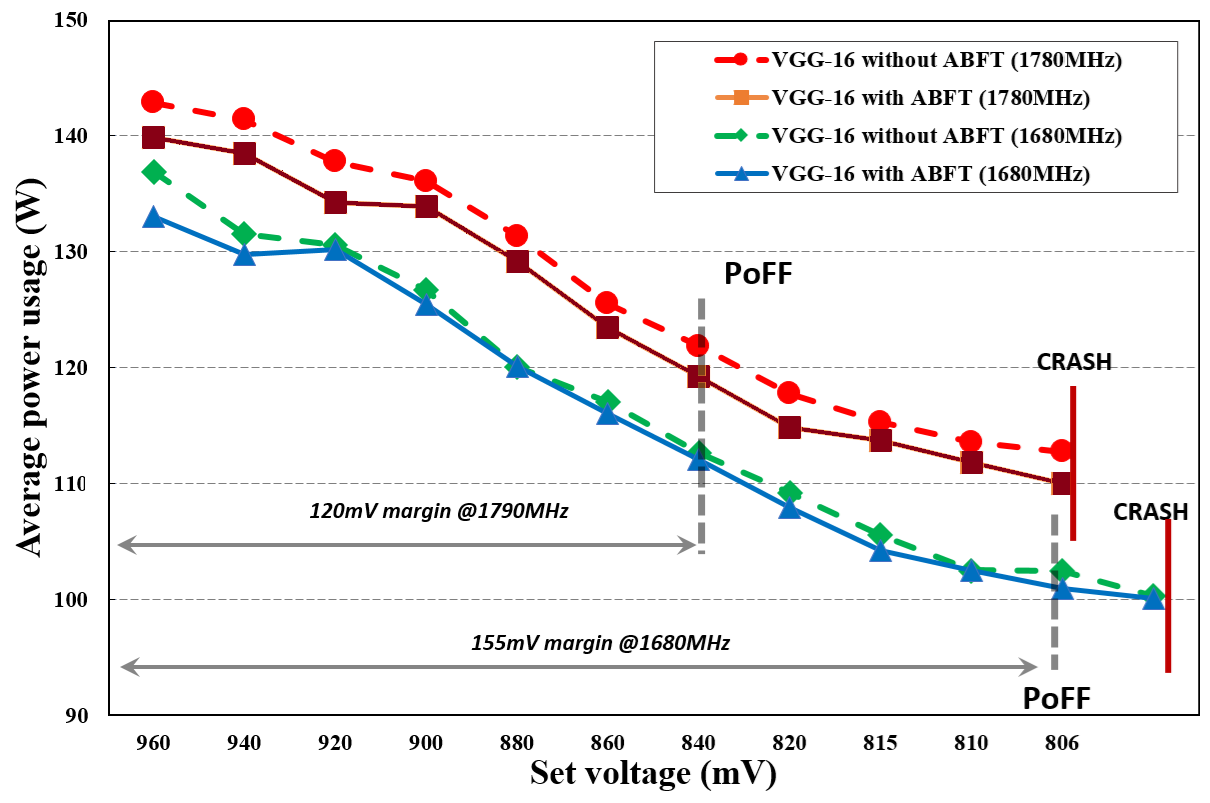}
  \vspace{-5pt}
  \caption{Power consumption with ABFT enabled and ABFT disabled for both VGG-16 at 1780~MHz and 1680~MHz clock frequencies with respect to voltage. The PoFF and crash points are marked for each.}  \label{powerConsumption}
  \end{center}
  \vspace{-5pt}
\end{figure}

\subsection{Power Saving }
A particularly interesting issue is the dynamic power savings of the GPU. As the voltage scaling quadratically reduces dynamic power, we kept highest utilization possible and scaled down the voltage till an error is observed by ABFT, i.e., PoFF is detected. Once PoFF is determined, one could stop reducing voltage; however, for characterization purposes we further reduced voltage down to the crash point. At the same time, the actual errors and impact of them on accuracy loss of the DNN model was measured. The bar, i.e., error threshold, for reporting using ABFT was set high which results in the higher error rate reported by ABFT compared to actual error rate. One could set a more relaxed error threshold, adding few more milliwatts of power savings. Figure~\ref{powerConsumption} depicts power savings trends while Fig.~\ref{errors} plots the error rate and accuracy loss for different voltages. Notice, the PoFF and crash points changes for different clock frequencies. 

We initially implemented ABFT on LeNet to assess its viability. However, due to the small model size of LeNet and the inherent overhead scaling of ABFT, which is $(1/N)$, where $N$, represents the problem size, we found that ABFT is not well-suited for very small deep neural networks (DNNs). While the preliminary results demonstrated notable power savings, the observed 7\% overhead could be misleading and may lead to incorrect generalizations about ABFT's efficiency. In contrast, VGG-16, with its significantly larger size, exhibited much lower relative overheads, making it a more appropriate candidate for evaluating ABFT's impact on larger DNNs. Hence, LeNet results were not plotted. Nonetheles, those are reported here. The power drops from around 30W down to 25W at 812mV (PoFF for LeNet) and 22W at around 805mV and $f_{clk}=1780MHz$.

Measurements results for power are averaged out of 120 inference experiments per operating points taken within 0.5s intervals. Table ~\ref{tab:infTimes} shows the power dissipation differences of the GPU at nominal voltage $V_{nominal}$ and three PoFF voltages $V_{min}$ and respective clock rates along with calculated energy overheads.

\begin{table}[t!]
\small
\centering
\caption{Power savings and energy overheads of ABFT-enabled VGG-16 at different clock frequencies.}
\vspace{-5pt}
\label{tab:power_savings}
\renewcommand{\arraystretch}{1.3} 
\setlength{\tabcolsep}{10pt} 
\begin{tabular}{lcccc}
\toprule
\textbf{Parameter} & \textbf{1820 MHz} & \textbf{1780 MHz} & \textbf{1680 MHz} \\
\midrule
\textbf{P($W$) @Nominal} & $141$ & $142$ & $137$ \\
\midrule
\textbf{$V_{min}$ (mV)} & $850$ & $835$ & $800$ \\
\textbf{P($W$) @$V_{min}$} & $116$ & $110$ & $107$ \\
\textbf{Energy/Inference ($J$)} &$ 20.7$ & $19.9 $&$ 19.0 $\\
\textbf{Energy Savings ($\%$)} &$ 18\% $& $21\%$ &$ 25\%$ \\
\textbf{Energy Overhead ($\%$)} & $3.9\%$ & $1.0\%$ &$ 3.0\%$\\
\bottomrule
\end{tabular}
\label{tab:infTimes}
\end{table}

\vspace{-7pt}

\begin{figure}[t!]
\small
  \begin{center}
  \label{errors}
\includegraphics[width=4in,height=2.55in]{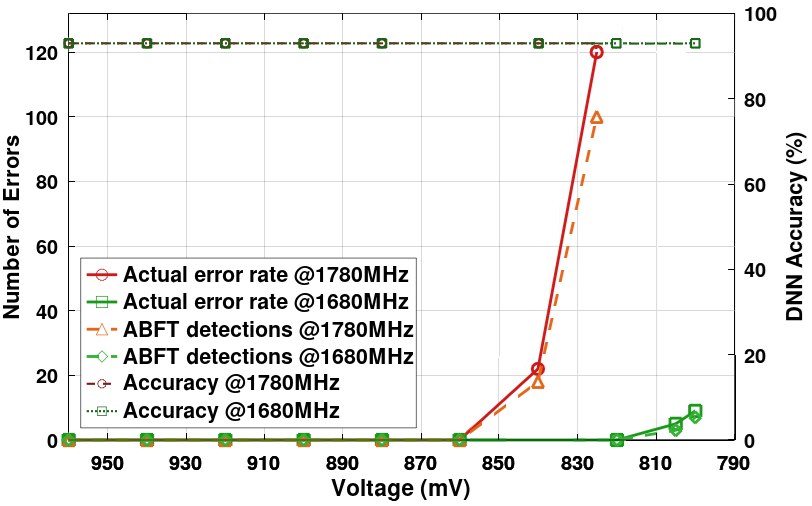}
  \vspace{-5pt}
  \caption{The voltage is reduced from detail down to the crash point. The ABFT detect errors in computations. Notice no accuracy loss is observed despite ABFT detections which is due to inherent fault tolerance of DNNs.}\label{errors}
  \end{center}
  \vspace{-15pt}
\end{figure}
\subsection{Error Coverage }

Figure~\ref{errors}, depicts the errors appearing in computations, i.e., actual error rate, error rate from ABFT detections and overall prediction accuracy of the model for a selected test set. Since errors propagate from one layer to another, comparing ABFT error detection rate against actual errors per each operations is misleading. Experiments on single operation, shows very high (close to 100\%) computational detection rate for ABFT, similiar to results on in \cite{safarpour2021high}. Errors were reported if difference between the GPU accelerated outputs and the reference was over
1e-13. The error threshold can be made slightly tighter to ensure more reliability, however, that would result in false positives being detected constantly, even at stock voltage and frequency settings. 

Notably, due to inherent fault-tolerance of DNNs a few hundred errors at vicinity of PoFF does not impact the accuracy, as it is observed in Fig.~\ref{errors}. Nonetheless, since ABFT detects the computational errors early enough, we do not need to rely on fault-tolerance feature of DNNs for ensuring reliability.

As expected, we observed that the errors appear in linear layers significantly before being detected in the non-linear ones. We speculate the possible explanation is that the circuit delay paths for non-linear operations are much shorter than those of floating-point multipliers  used for arithmetic computations.

\subsection{Overheads}

The power consumption at all voltages with ABFT enabled is slightly lower (around 1\%) than when it is disabled, meaning ABFT overheads manifest in inference time rather than power. That is likely due to ABFT related calculations imposing more idle periods on processing units. As Table ~\ref{tab:infTimes} shows, this translates into slightly longer inference time. Overall energy overhead estimated by comparing the energy per inferences between ABFT enabled and disabled runs. Table~\ref{TimesOverheads} provides a summary of overheads in terms of inference time.

Naive integration of ABFT into the DNN model results in large overheads, however, with small tweaks and code optimization the overhead share becomes insignificant, e.g.,~$<3.9\%$ for VGG-16. Considering VGG-16 with default clock frequency and voltage (at 1780~MHz and 960~mV), the energy overhead of ABFT is as low as 1\%. With same frequency but voltage set to  880~mV, the energy overhead  is almost the same, i.e.,~1.3\%. Setting the frequency to the lowest, i.e., 1680~MHz, and at the lowest stable around 805~mV, the energy overhead of ABFT enabled is slightly higher, i.e., 1.9\%. This is a small penalty for saving almost a quarter of the energy of each inference.

\begin{table}[t!]
\small
\centering
\caption{Inference times \mbox{VGG-16} at different clock frequencies.}
\vspace{-5pt}
\label{TimesOverheads}
\renewcommand{\arraystretch}{1.3} 
\setlength{\tabcolsep}{5pt} 
\begin{tabular}{lccc}
\toprule
\textbf{Configuration} & \textbf{VGG-16} & \textbf{VGG-16} & \textbf{VGG-16} \\
 & \textbf{@ 1820 MHz} & \textbf{@ 1780 MHz} & \textbf{@ 1680 MHz} \\
\midrule
\textbf{ABFT disabled}~($ms)$ & 171.90 & 175.19 & 183.42 \\
\textbf{ABFT enabled}~($ms$) & 178.08 & 181.36 & 189.86 \\
\midrule
\textbf{Time overhead (\%)} &  & $\approx3.5\%$ &     \\
\bottomrule
\end{tabular}
    \label{Overheads}
\end{table}

\section{Related Works and Discussion}

In \cite{gundi2020effort}, a Tensor Processing Unit (TPU) architecture, named "EFFORT", is proposed. The TPU is equipped with timing error detection circuits within critical paths of Multiply and Accumulate (MAC) units to enable reduced voltage operations. EFFORT achieves high performance and energy efficiency by employing TED systems and in-situ clock gating techniques to address. To minimize hardware overheads, not all critical paths can be equipped with TEDs and hence a 2\% average accuracy drop across various DNN datasets was reported.

Similar to EFFORT, in \cite{fan2022design} a lightweight error detection circuit is introduced to detect timing errors from reduced voltage operations. However, circuit level techniques substantially complicate design and test process and are not applicable to already fabricated chips. Notice, that our proposed solution does not compromise the accuracy of DNN. MATIC \cite{kim2018matic} combines algorithm-level and circuit-level techniques to enable efficient low-voltage operation of neural accelerators. No online error detection is utilized, instead, MATIC addresses the challenge of timing errors based on speculative fault model and error-aware training. Hence, once the fault model no longer is applicable, the accuracy can not be ensured. 

One of the earliest works on algorithmic approaches for for enabling reduced voltage operation was introduced in \cite{zamani2019greenmm}. However, they limit the application only to general matrix multiplication. We solved this limitation by developing ABFT enabled CNN models to demonstrate possibility for improved  energy efficiency and reliable reduced voltage acceleration of CNNs.

This work can be extended for training as well. The solution was only showcased on a GPU, while a similar approach can be applied to other DNN accelerators, e.g., Google TPUs. Due to software based nature of this work, Shavette, is more cost efficient compared to hardware based methods, e.g., Razor \cite{ernst2003razor}. A summary of comparison with example similiar works is presented in Table~\ref{Table_comparisons}.

Although, we resorted to implementing the models from scratch in OpenCL, the necessary modifications can be made to the DNN automatically using development tools such as TensorFlow or Keras \cite{zhao2020ft} when better software support from vendor is available. Nonetheless, this work demonstrates the possibility of a low-cost software-based voltage scaling solution for DNN acceleration.

 \begin{table}[bt!]
\small
\centering
 \caption{COMPARISON WITH OTHER WORKS.}
\vspace{-5pt}
\begin{tabular}{lcccc}
\hline
Ref.     & \multicolumn{1}{l}{Application} &   \multicolumn{1}{l}{~~~~~~Detection} & \multicolumn{1}{l}{Overheads} & \multicolumn{1}{l}{Coverage} \\ \hline
 \cite{gundi2020effort}& General        & Online HW (TED)       & 3.5\%                  & N/A           \\  
    \cite{zamani2019greenmm}   & Matrix          & Online Algorithmic (ABFT)   & $\approx$1\%-10\%     &$\approx$100\%    \\
 \cite{safarpour2022lofft}   & Fourier          & Online Algorithmic (Parseval's Identity)    & $\approx$8\%    &$\approx$100\%                         \\

\cite{maragos2021pvt}      & General       & Offline HW (Calib.)    & 1.6\%                   & $\approx$95\%             \\

\cite{jiang2022fodm}       & General      & Offline HW (Calib.)    & 0.4\%-1.5\%             & \textgreater{}98.5\%           \\ 
\cite{zhang2018thundervolt}& Matrix        & Online HW (TED)       & 30\%                  & N/A            \\  

\hline
Shavette                   & DNN          & Online Algorithmic (ABFT) & $\approx$3-7\%       & $\approx$100\%       \\ \hline
\end{tabular}

\vspace{-10pt}
\label{Table_comparisons}
\end{table}
One of the issues that was verified in the context of this paper is that when the voltage is reduced the control path fails long after the data path. Nonetheless, for future works, the control path can be aggressively pipelined or be operated at slightly higher voltage such that it is ensured that data path always fails before control path. 

The proposed solution in this letter can be further investigated for enabling more aggressive voltage scaling schemes, e.g., Near Threshold Computing (NTC), which is promising up to 10x improvement in energy efficiency \cite{ernst2003razor}.


\section{Conclusion}
 In computations of DNN models, the voltage margins of the accelerator can be safely optimized as DNN models mostly comprised of operations in which ABFT schemes can be integrated into. This enables timely detection of voltage induced errors with minimal overheads. The time and energy overheads from ABFT are low, e.g., around 3.5\% and 1.1\% respectively. This work demonstrated that nearly a quarter of energy could be saved without sacrificing accuracy of the model. The results indicate  a promising avenue for enhancing the energy efficiency of AI accelerators, without expensive HW techniques or having to compromise reliability. 

\section*{Acknowledgment}
This work was supported by 6G Flagship (Grant Number 369116) funded by the Research Council of Finland.


\begin{thebibliography}{10}

\bibitem{tarkoma2023ai}
S.~Tarkoma, R.~Morabito, and J.~Sauvola, ``Ai-native interconnect framework for integration of large language model technologies in 6g systems,'' {\em arXiv preprint arXiv:2311.05842}, 2023.

\bibitem{krishna2022global}
C.~Krishna, ``Global voltage scaling across multiple cores for real-time workloads,'' {\em IEEE Embedded Systems Letters}, vol.~14, no.~3, pp.~159--162, 2022.

\bibitem{ernst2003razor}
D.~Ernst, N.~S. Kim, S.~Das, S.~Pant, R.~Rao, T.~Pham, C.~Ziesler, D.~Blaauw, T.~Austin, K.~Flautner, {\em et~al.}, ``Razor: A low-power pipeline based on circuit-level timing speculation,'' in {\em Proceedings. 36th Annual IEEE/ACM International Symposium on Microarchitecture, 2003. MICRO-36.}, pp.~7--18, IEEE, 2003.

\bibitem{safarpour2021high}
M.~Safarpour, L.~Xun, G.~V. Merrett, and O.~Silv{\'e}n, ``A high-level approach for energy efficiency improvement of fpgas by voltage trimming,'' {\em IEEE Transactions on Computer-Aided Design of Integrated Circuits and Systems}, vol.~41, no.~10, pp.~3548--3552, 2021.

\bibitem{thomas2016emergpu}
R.~Thomas, N.~Sedaghati, and R.~Teodorescu, ``Emergpu: Understanding and mitigating resonance-induced voltage noise in gpu architectures,'' in {\em 2016 IEEE International Symposium on Performance Analysis of Systems and Software (ISPASS)}, pp.~79--89, IEEE, 2016.

\bibitem{hiienkari20200}
M.~Hiienkari, N.~Gupta, J.~Teittinen, J.~Simonsson, M.~Turnquist, J.~Eriksson, R.~Anttila, O.~Myllynen, H.~R{\"a}m{\"a}kk{\"o}, S.~M{\"a}kikyr{\"o}, {\em et~al.}, ``A 0.4-0.9 v, 2.87 pj/cycle near-threshold arm cortex-m3 cpu with in-situ monitoring and adaptive-logic scan,'' in {\em 2020 IEEE Symposium in Low-Power and High-Speed Chips (COOL CHIPS)}, pp.~1--3, IEEE, 2020.

\bibitem{kim2016adaptive}
J.~Kim, G.~Lee, K.~Choi, Y.~Kim, W.~Kim, K.~Do, and J.~Choi, ``Adaptive delay monitoring for wide voltage-range operation,'' in {\em 2016 Design, Automation \& Test in Europe Conference \& Exhibition (DATE)}, pp.~511--516, IEEE, 2016.

\bibitem{do2023evaluating}
D.~V.~C. do~Nascimento, K.~Georgiou, K.~I. Eder, and S.~Xavier-de Souza, ``Evaluating the effects of reducing voltage margins for energy-efficient operation of mpsocs,'' {\em IEEE Embedded Systems Letters}, 2023.

\bibitem{safarpour2021algorithm}
M.~Safarpour, R.~Inanlou, and O.~Silv{\'e}n, ``Algorithm level error detection in low voltage systolic array,'' {\em IEEE Transactions on Circuits and Systems II: Express Briefs}, vol.~69, no.~2, pp.~569--573, 2021.

\bibitem{huang1984algorithm}
K.-H. Huang and J.~A. Abraham, ``Algorithm-based fault tolerance for matrix operations,'' {\em IEEE transactions on computers}, vol.~100, no.~6, pp.~518--528, 1984.

\bibitem{ensieh2025scissors}
A.~Ensieh, M.~Safarpour, C.~Wulf, O.~Silven, and D.~Gohringer, ``Scissors: System level error detection for enabling near-threshold operating systolic arrays,'' {\em arXiv preprint arXiv:2405.24015}, 2025.

\bibitem{marty2020safe}
T.~Marty, T.~Yuki, and S.~Derrien, ``Safe overclocking for cnn accelerators through algorithm-level error detection,'' {\em IEEE Transactions on Computer-Aided Design of Integrated Circuits and Systems}, vol.~39, no.~12, pp.~4777--4790, 2020.

\bibitem{zhao2020ft}
K.~Zhao, S.~Di, S.~Li, X.~Liang, Y.~Zhai, J.~Chen, K.~Ouyang, F.~Cappello, and Z.~Chen, ``Ft-cnn: Algorithm-based fault tolerance for convolutional neural networks,'' {\em IEEE Transactions on Parallel and Distributed Systems}, vol.~32, no.~7, pp.~1677--1689, 2020.

\bibitem{lecun1998gradient}
Y.~LeCun, L.~Bottou, Y.~Bengio, and P.~Haffner, ``Gradient-based learning applied to document recognition,'' {\em Proceedings of the IEEE}, vol.~86, no.~11, pp.~2278--2324, 1998.

\bibitem{nia2007generalized}
A.~M. Nia and K.~Mohammadi, ``A generalized abft technique using a fault tolerant neural network,'' {\em Journal of Circuits, Systems, and Computers}, vol.~16, no.~03, pp.~337--356, 2007.

\bibitem{amdgpu_clocks}
sibradzic, ``amdgpu-clocks.'' \url{https://github.com/sibradzic/amdgpu-clocks}.
\newblock Accessed: 2025-03-18.

\bibitem{MikaelsThesis}
M.~Rinkinen, ``Low voltage gpu-based ai accelerator,'' Master's thesis, University of Oulu, 2024.

\bibitem{gundi2020effort}
N.~D. Gundi, T.~Shabanian, P.~Basu, P.~Pandey, S.~Roy, K.~Chakraborty, and Z.~Zhang, ``Effort: Enhancing energy efficiency and error resilience of a near-threshold tensor processing unit,'' in {\em 25th Asia and South Pacific Design Automation Conference}, pp.~241--246, IEEE, 2020.

\bibitem{fan2022design}
X.~Fan, H.~Liu, H.~Li, S.~Lu, and J.~Han, ``Design of light-weight timing error detection and correction circuits for energy-efficient near-threshold voltage operation,'' {\em Electronics}, vol.~11, no.~18, p.~2879, 2022.

\bibitem{kim2018matic}
S.~Kim, P.~Howe, T.~Moreau, A.~Alaghi, L.~Ceze, and V.~Sathe, ``Matic: Learning around errors for efficient low-voltage neural network accelerators,'' in {\em 2018 Design, Automation \& Test in Europe Conference \& Exhibition (DATE)}, pp.~1--6, IEEE, 2018.

\bibitem{zamani2019greenmm}
H.~Zamani, Y.~Liu, D.~Tripathy, L.~Bhuyan, and Z.~Chen, ``Greenmm: energy efficient gpu matrix multiplication through undervolting,'' in {\em Proceedings of the ACM International Conference on Supercomputing}, pp.~308--318, 2019.

\bibitem{safarpour2022lofft}
M.~Safarpour and O.~Silv{\'e}n, ``Lofft: Low-voltage fft using lightweight fault detection for energy efficiency,'' {\em IEEE Embedded Systems Letters}, 2022.

\bibitem{maragos2021pvt}
K.~Maragos, G.~Lentaris, and D.~Soudris, ``A pvt-aware voltage scaling method for energy efficient fpgas,'' in {\em 2021 IEEE International Symposium on Circuits and Systems (ISCAS)}, pp.~1--5, IEEE, 2021.

\bibitem{jiang2022fodm}
W.~Jiang, H.~Yu, H.~Zhang, Y.~Shu, R.~Li, J.~Chen, and Y.~Ha, ``Fodm: A framework for accurate online delay measurement supporting all timing paths in fpga,'' {\em IEEE Transactions on Very Large Scale Integration (VLSI) Systems}, vol.~30, no.~4, pp.~502--514, 2022.

\bibitem{zhang2018thundervolt}
J.~Zhang, K.~Rangineni, Z.~Ghodsi, and S.~Garg, ``Thundervolt: enabling aggressive voltage underscaling and timing error resilience for energy efficient deep learning accelerators,'' in {\em Proceedings of the 55th Annual Design Automation Conference}, pp.~1--6, 2018.

\end{thebibliography}

\end{document}